\documentclass[10pt]{iopart}
\usepackage[utf8]{inputenc}
\usepackage{tabularx}
\usepackage{array}
\usepackage{float}
\usepackage{graphicx}
\usepackage{longtable}
\usepackage{textcomp}
\usepackage{color}
\usepackage{cite}
\usepackage{gensymb}

\newcommand{\modif}[1]{\textcolor{black}{#1}}

\begin{document}

\title[Anomalous depinning of magnetic domain walls in Co$_3$Sn$_2$S$_2$]{Anomalous depinning of magnetic domain walls within the ferromagnetic phase of the Weyl semimetal Co$_3$Sn$_2$S$_2$}

\author{Zihao Shen}
\address{Department of Physics and Astronomy, University of California, Davis, California 95616, USA}
\author{X.D. Zhu}
\address{Department of Physics and Astronomy, University of California, Davis, California 95616, USA}
\address{Department of Optical Sciences and Engineering, Fudan University, Shanghai 200045, China}
\ead{xdzhu@physics.ucdavis.edu}
\author{Rahim R. Ullah}
\address{Department of Physics and Astronomy, University of California, Davis, California 95616, USA}
\author{Peter Klavins}
\address{Department of Physics and Astronomy, University of California, Davis, California 95616, USA}
\author{Valentin Taufour}
\address{Department of Physics and Astronomy, University of California, Davis, California 95616, USA}
\ead{vtaufour@ucdavis.edu}

\begin{abstract}

We report bulk magnetization measurements and spatially resolved
measurements of magnetic domains in
$\mathrm{Co}_{3}\mathrm{Sn}_2\mathrm{S}_2$ single crystals. The results indicate that a previously reported magnetic anomaly around
$130$\,K is due to an anomalous domain wall depinning upon cooling. Our measurements also reveal a hysteresis between field-cooled-cooling (FCC) and field-cooled-warming (FCW) magnetization curves acquired under a constant magnetic field 
below 300\,Oe. This observation rules out the possibility that the anomaly stems from a second-order phase transition. Our results further suggest that changes in the shape of hysteresis loops from $5$\,K to $170$\,K are caused by an unusual temperature-dependent domain nucleation field that changes sign around $130$\,K. The Kerr rotation images of the magnetic domains confirm that the domain walls depin between $120$\,K and $140$\,K.

\end{abstract}

\maketitle

\ioptwocol

\section{Introduction}

Weyl Semimetals (WSM) are material systems with non-degenerate bands that cross each other near the Fermi level when inversion symmetry or time-reversal symmetry is broken~\cite{Wan,Burkov,Armitage}. As one of the most interesting examples of WSM~\cite{Liu}, Co$_3$Sn$_2$S$_2$ is a ferromagnetic material with a Curie temperature of $T_\mathrm{C}=174$\,K~\cite{Yan}. The topological features of this material lead to a large anomalous Hall conductivity~\cite{Wang}, an anomalous Nernst Hall effect~\cite{Yang}, a negative magnetoresistance due to a chiral anomaly~\cite{Liu}, and an unusual breaking-down of Ohm's law~\cite{Nagpal}. In addition, an unusually large magneto-optic effect has also been observed~\cite{Okamura}.

Co$_3$Sn$_2$S$_2$ crystallizes in a hexagonal lattice with triangular S and Sn layers interspersed between the kagome lattice planes of Co. The magnetic structure of this compound is still under investigation due to an anomaly in its magnetization occurring around $130$\,K upon cooling~\cite{Kassem_M,kassemunconventional,Q_ZhangPRL,Soh,shin,Hu,Shen}. Various interpretations have been proposed to explain the anomaly. Wu et al. proposed a hidden magnetic phase in Co$_3$Sn$_2$S$_2$ as revealed in AC susceptibility measurements~\cite{Wu}. Guguchia et al. suggested an antiferromagnetic phase (AFM) in this compound based on two different muon precessions below $T_\mathrm{C}$~\cite{Guguchia}. Meanwhile, other groups interpreted the anomaly as a spin glass transition~\cite{Lachman,Q_Zhang}, a domain wall transition from linear domain walls to elliptical domain walls~\cite{Lee}, and an enhanced domain wall motion observed by magnetic force microscopy~\cite{Howlader}. Some of these proposals, however, have been challenged by experiments. For example, no strong evidence for an AFM phase was found in a neutron scattering experiment~\cite{Soh}. 

In this context, we report bulk magnetization measurements on Co$_3$Sn$_2$S$_2$ single crystals in external magnetic fields applied along the $c$ axis, where we find a depinning of magnetic domain walls around $130$\,K upon cooling and a hysteresis between field-cooled-cooling (FCC) and field-cooled-warming (FCW) magnetization at low fields. We also present magneto-optic Kerr effect (MOKE) images of magnetic domains in Co$_3$Sn$_2$S$_2$ from $120$ to $140$\,K obtained under an external magnetic field of $94$\,Oe applied along the $c$ axis. \modif{From the MOKE images,} we extract fractions of single domains and domain walls near the magnetic anomaly around $130$\,K. While we find no strong evidence of sudden domain reorganization or domain size changes, we do find a change in domain fractions consistent with the partial depinning of magnetic domains. Our results indicate that the magnetic anomaly can be explained by a domain wall depinning effect. The origin of the domain wall depinning effect remains to be further investigated.

\section{Methods}

\subsection{Single Crystal Growth}
Single crystals of Co$_3$Sn$_2$S$_2$ were synthesized by solution growth~\cite{Nagpal,Lin,Kassem}. A ternary mixture with initial composition Co$_{12}$S$_8$Sn$_{80}$ was first heated to $400$\degree C over two hours and held there for another two hours. It was then heated to $1050$\degree C over six hours and held there for $10$~hours, followed by a slow, $90$ hour cool down to $740$\degree C. The remaining molten flux was removed by centrifugation. Shiny hexagonal crystals were obtained. The powder x-ray diffraction data refines to a hexagonal unit cell with $a=5.3641(8)$\,\AA \,and $c=13.1724$\,\AA, which are consistent with previously reported values~\cite{Vaqueiro}.

\subsection{Magnetization measurements and magnetic domain imaging}
Magnetization measurements of Co$_3$Sn$_2$S$_2$ single crystals were performed in a Quantum Design MPMS SQUID up to $7$\,T with the magnetic field parallel to the $c$ axis. Magnetization as a function of temperature was measured under various applied fields in both field-cooled-cooling (FCC) and field-cooled-warming (FCW) conditions. Hysteresis loops were measured by first cooling down to the desired temperature in zero applied field and then sweeping the field. Before each measurement, we followed a demagnetization procedure to minimize the remanent field in the magnet and in the sample chamber by systematically oscillating the magnet to zero field at room temperature.

Polar MOKE images are acquired from as-grown Co$_3$Sn$_2$S$_2$(0001) single crystal samples, using a normal-incidence Sagnac interferometric scanning microscope ~\cite{Zhu}. The microscope measures Kerr rotation as opposed to Kerr ellipticity~\cite{Okamura}. It has a spatial resolution of 0.85\,$\mu$m and a sensitivity of $4.4$\,$\mu$rad. The single crystal sample is mounted on a cold finger inside a cryostat. The sample is optically accessible from outside through an optical window. The temperature of the sample is variable from $300$\,K to $5$\,K. We use a permanent magnet placed outside of the cryostat to produce a variable magnetic field of up to 100\,Oe (0.01\,T) along the $c$ axis (the surface normal) of the sample. For this experiment, the field is held constant at 94\,Oe, measured by a Hall effect sensor.

\section{Results and Discussion}

\subsection{Magnetization measurements}
\begin{figure}[h]
\centering
    \includegraphics[width=0.48\textwidth]{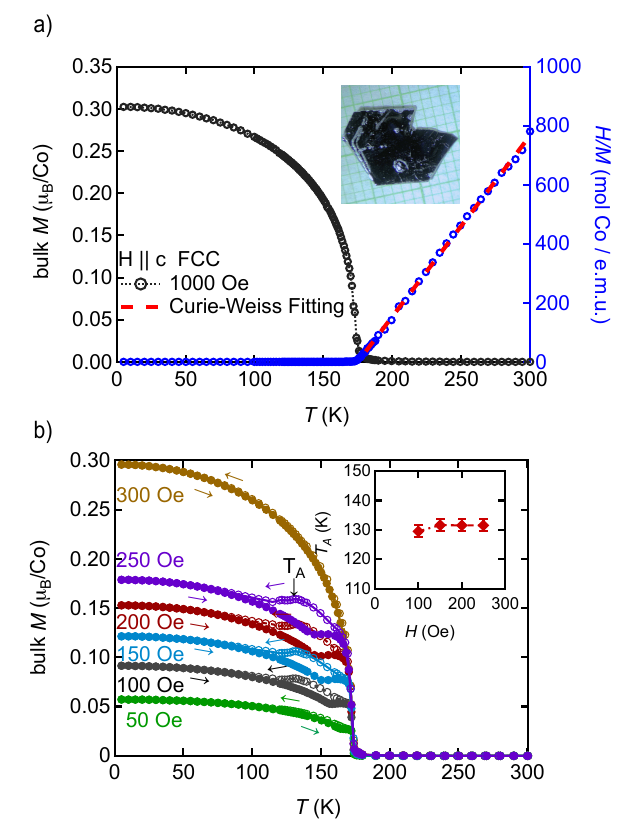}
    \caption{a) Magnetization as a function of temperature measured by FCC in a $1000$\,Oe applied field parallel to the $c$ axis. The blue curve shows a Curie-Weiss fitting with an effective moment $1.14$\,$\mu_B$ per Co. The inset shows a Co$_3$Sn$_2$S$_2$ single crystal grown from solution growth. b) FCC and FCW curves show a hysteresis between $100$\,K and $T_\mathrm{C}=174$\,K. $T_A$ is the temperature of the magnetic anomaly where the FCC magnetization reaches a local maximum followed by a downward kink. The inset shows the field dependence of $T_{A}$.}
    \label{fig:Magnetization}
\end{figure}

The FCC curve with a magnetic field of $1000$\,Oe applied parallel to the $c$ axis is shown in Fig.~\ref{fig:Magnetization}, and illustrates the ferromagnetic-paramagnetic transition. The inverse susceptibility is fit to the Curie-Weiss law and we obtain an effective moment of $1.14$\,$\mu_B$ per Co and a Curie-Weiss temperature of $\theta_\mathrm{CW}=175$\,K, which are consistent with previously published work~\cite{Lin,Vaqueiro,Schnelle,Liu}. There is no difference in the FCC and FCW magnetization at fields above $300$\,Oe. If the applied field is below $300$\,Oe, however, two anomalies emerge. First, the FCC and FCW curves start to deviate near $T_\mathrm{C}$. Second, a local maximum arises at $T_\mathrm{A}$ $\approx$ 130\,K in the FCC data, which is also reported in other studies~\cite{Guguchia,Kassem_M,kassemunconventional,Q_ZhangPRL,Soh,shin,Hu,Shen}. The local maximum disappears above 300\,Oe and below 50\,Oe, while a small hysteresis between FCC and FCW still remains at 50\,Oe. The field dependence of $T_\mathrm{A}$ is relatively small, as shown in the inset of Fig.~\ref{fig:Magnetization}(b). The fact that the anomaly disappears above $300$\,Oe, where the magnetization is saturated, suggests its origin is related to magnetic domains. A local minimum also emerges in the FCW magnetization albeit at higher temperature. The hysteresis between FCC and FCW magnetizations indicates it is unlikely that the local maximum and minimum in the FCC and FCW magnetization are due to a second order phase transition, which should not depend on magnetic history.

\begin{figure}[h]
\centering
    \includegraphics[width=0.48\textwidth]{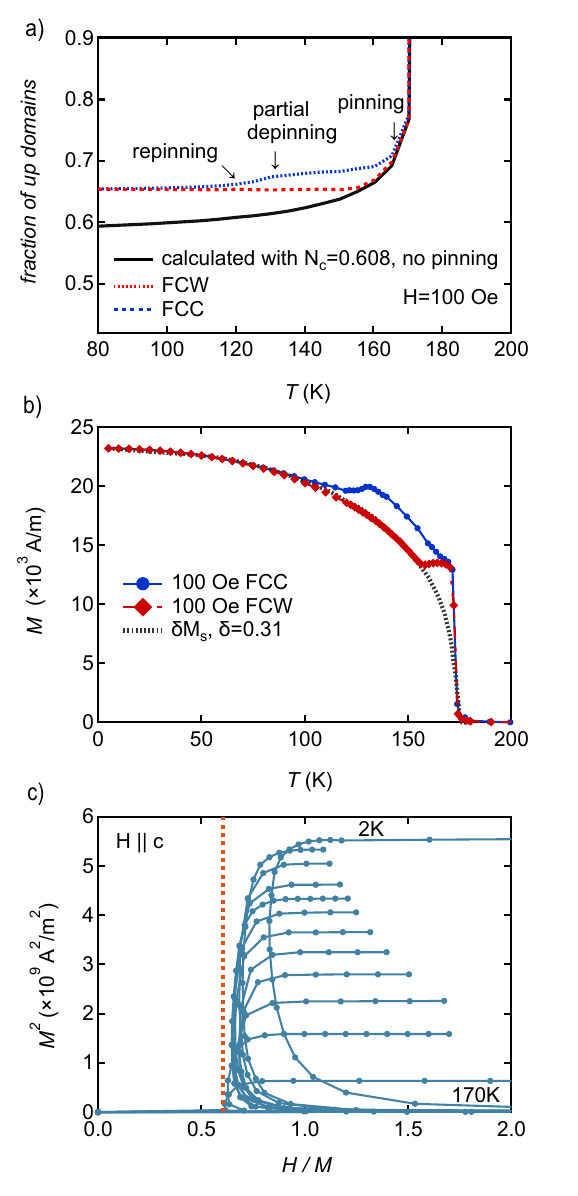}
    \caption{a) Temperature dependence of the fraction of up domains extracted from FCC (blue curve) and FCW (red curve) magnetization data at $H = 100$\,Oe compared to our model based on demagnetization theory with $N_c = 0.608$ (black curve). b) Comparison between FCC and FCW magnetization at 100\,Oe and $\delta M_\mathrm{s}$ with $\delta = 0.31$. \modif{c) $M^{2}$ versus $H/M$ at various temperatures. The intercept on the x-axis (orange line) reveals a demagnetization factor $N_\mathrm{c}$ of $0.608$.}}  
    \label{fig:fraction_up_FCC_FCW}
\end{figure}

\begin{figure*}[!htb]
\centering
    \includegraphics[width=1\textwidth]{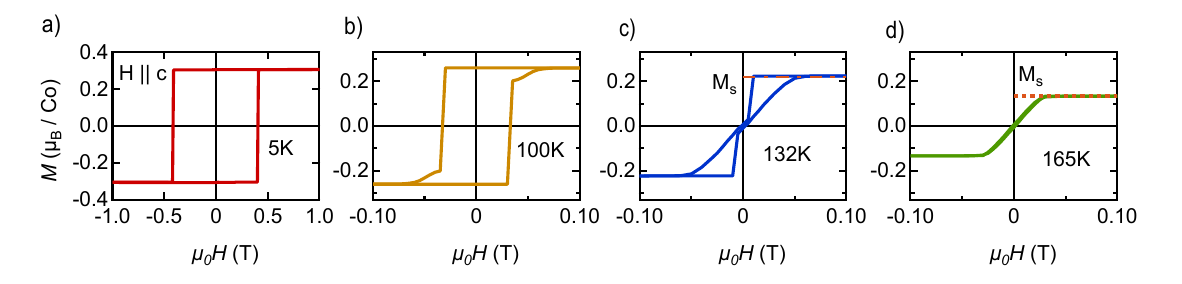}
    \caption{Hysteresis loops with the applied field up to $7$\,T parallel to the $c$ axis at various temperatures. Only a selection of temperatures are shown. The hysteresis gradually changes from rectangle to bi-triangular shape. \modif{Spontaneous magnetization $M_\mathrm{s}$ is obtained as shown in (c) and (d).}} 
    \label{fig:hysteresis}
\end{figure*}

We will compare the fraction of magnetic domains assuming no pinning with the fraction extracted from our experimental magnetization where pinning effects are inevitable. In the absence of domain wall pinning, the demagnetising field $H_{d}=N_{c}M$ perfectly counteracts the applied field, $H_\mathrm{applied}$, as long as $H_\mathrm{applied}<N_{c}M_\mathrm{s}$. \modif{$N_c$ is the demagnetization factor along the $c$-direction. Because the $c$ axis is the easy magnetization axis, $N_c$ can be measured experimentally from a plot of $M^2$ versus $H_\mathrm{applied}/M$ for the $M(H)$ data along the easy axis of magnetization as shown in Fig.~\ref{fig:fraction_up_FCC_FCW}(c). The x-intercept gives the value of the demagnetization factor along the easy magnetization axis~\cite{lamichhane2016study,ArrottPRB}. Due to the existence of domain pinning, the curves at low temperatures tend to shift to the right. Therefore, we used the x-intercept at $170$\,K, where domain pinning is negligible, as the demagnetization factor and obtained $N_c = 0.608$. $M_\mathrm{s}$ is the spontaneous magnetization, which is obtained from the extrapolation to zero field of $M(H)$ curves, as shown in Fig.~\ref{fig:hysteresis} (c) and (d).} The magnetization of the sample can be related to the fraction of up domains $n_{\uparrow}$, where $M = (2n_{\uparrow}-1)M_\mathrm{s}$. From $H_\mathrm{applied}=N_{c}M$, we can determine $n_{\uparrow}$ as follows:

\begin{equation}\label{eq:domainfraction}
    n_{\uparrow} = \frac{1}{2M_\mathrm{s}}\frac{H_\mathrm{applied}}{N_{c}}+ \frac{1}{2}
\end{equation}
The applied field, $H_\mathrm{applied}$, and the demagnetization factor, $N_{c}$, are temperature independent. Equation~\ref{eq:domainfraction} gives us the result that, without pinning, the fraction of up domains should decrease as temperature decreases because $M_\mathrm{s}$ increases.

In Fig.~\ref{fig:fraction_up_FCC_FCW}(a) we compare the calculated fraction of magnetically aligned (up) domains with the value deduced from the experimental magnetization. \modif{The blue and red curves are the temperature dependence of the up-domain fraction extracted from the FCC and FCW magnetization respectively at $100$\,Oe, with the following formula:}

\begin{equation}\label{eq:updomainfraction}
    n_{\uparrow} = \frac{M}{2M_\mathrm{s}}+ \frac{1}{2}.
\end{equation}
The black curve is the fraction of up-domain calculated from a demagnetization theory~\cite{Tremolet} in Eq.~\ref{eq:domainfraction}. We can see that domain wall pinning keeps the up-domain fraction above the theoretical value for all but the temperatures near $T_C$. While cooling down below $T_C$, the system encounters a domain-wall pinning effect where the blue curve begins to deviate from the black curve. This domain-wall pinning causes the magnetization to increase nearly proportionally to $M_s$ as shown by the blue curve in Fig.~\ref{fig:fraction_up_FCC_FCW}(b). Below around $132$\,K, the system enters a partial depinning region where the up-domain fraction decreases towards its theoretical no pinning value. The magnetization consequently decreases, which causes a local maximum near $132$\,K. Upon further cooling, there is a competition between this anomalous depinning effect and the reduced thermal energy which tends to favor pinning. The reduced thermal energy eventually dominates around $120$\,K and domain walls tend to pin again. Therefore, the magnetization tends to be proportional to $M_\mathrm{s}$, i.e. $M=\delta M_\mathrm{s}$ with $\delta=0.31$. The repinning is consistent with the fact that domain walls in Co$_3$Sn$_2$S$_2$ rarely move below 119\,K as reported by a Lorentz microscopy study~\cite{Sugawara}. Upon warming up, as shown by the red curve in Fig.~\ref{fig:fraction_up_FCC_FCW}, since domain walls are pinned, the domain fraction remains unchanged until around 155\,K, where the increased thermal energy allows more up domains to form. Due to the presence of domain-wall pinning, the total magnetization of the sample when cooled down and warmed up follows two different pathways, resulting in the hysteresis between the FCC and FCW magnetization curves.

\begin{figure}[h]
\centering
    \includegraphics[width=0.48\textwidth]{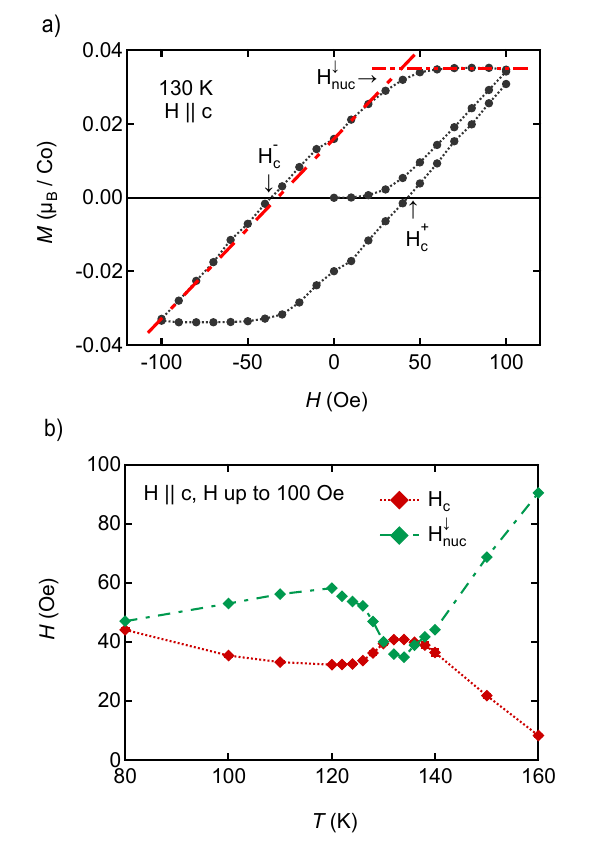}
    \caption{a) Hysteresis loop with applied field parallel to the $c$ axis up to $100$\,Oe. The arrow shows the coercive field $H_\mathrm{c}$. Nucleation field $H_\mathrm{nuc}^{\downarrow}$ is obtained by the intersection of red lines. b) $H_\mathrm{c}$ and $H_\mathrm{nuc}^{\downarrow}$, with maximum field up to $100$\,Oe, as a function of temperature.}  
    \label{fig:hys_Hc_100Oe}
\end{figure}

\begin{figure}[h]
\centering
    \includegraphics[width=0.45\textwidth]{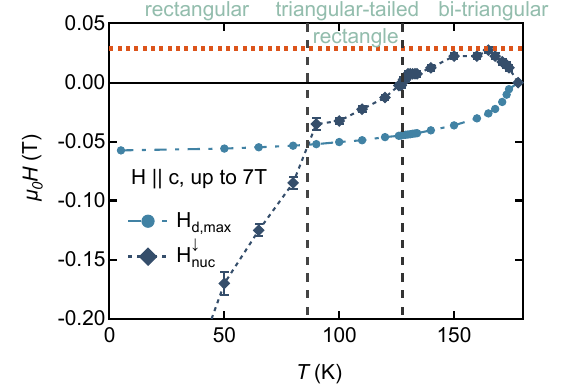}
    \caption{Temperature dependence of nucleation field and maximum demagnetization field. The red dashed line shows the $300$\,Oe field.} 
    \label{fig:Hnuc_T}
\end{figure}

For evidence of domain-wall depinning effects, we measured hysteresis loops with the applied field only up to $100$\,Oe at various temperatures. An example is shown in Fig.~\ref{fig:hys_Hc_100Oe}a. We note that domain structures depend on defects and the shape of a sample~\cite{Vial,Zeper}. For this reason, we use the same sample for all the bulk magnetization measurements presented in this article. The virgin curve at $130$\,K remains flat at first, then increases with a constant slope, which is the hallmark of a domain-wall pinning-type virgin curve\cite{Tremolet}. The magnetization of the field-increasing part of the hysteresis loop is smaller than that of the virgin curve, indicating that the domains encounter some pinning effect. Asymmetry between positive and negative coercive field emerges up to $160$\,K. The asymmetry can be attributed to the Exchange Bias (EB), which has been observed by other groups~\cite{Lachman,Noah}, or the remanent field during the measurement. At $130$\,K, the negative coercive field is $H_\mathrm{c}^{-}=-34.3(5)$\,Oe, while the positive coercive field is $H_\mathrm{c}^{+}=42.1(3)$\,Oe, and the actual coercive field, $H_{\mathrm{c}}$, is taken to be the average of the absolute value of the two. The nucleation field $H_\mathrm{nuc}^{\downarrow}$, below which the down domains start to form, was obtained as shown in Fig.~\ref{fig:hys_Hc_100Oe}(a). If there were no domain-wall pinning, the $M(H)$ curve should be a straight line. Therefore, the coercive field should always be zero and the nucleation field $H_\mathrm{nuc}^{\downarrow}$ should be the highest field of the hysteresis loop, i.e., $100$\,Oe. However, Fig.~\ref{fig:hys_Hc_100Oe}(b) reveals a local maximum of $H_\mathrm{c}$ and a local minimum of $H_\mathrm{nuc}^{\downarrow}$ at $132$\,K, below which $H_\mathrm{c}$ starts to decrease and $H_\mathrm{nuc}^{\downarrow}$ starts to increase, indicating a partial domain-wall depinning. The domain wall depinning around 130\,K is consistent with the high mobility of domain walls observed near 130\,K with a magnetic force microscope~\cite{Howlader}. Further below $120$\,K, domain-wall pinning strength increases again. $H_\mathrm{c}$ consequently increases and $H_\mathrm{nuc}^{\downarrow}$ decreases. We note that the local maximum disappears at fields above $300$\,Oe. As shown in Fig~\ref{fig:Hnuc_T}, $300$\,Oe is the largest value of the nucleation field when this sample is fully magnetized. Therefore, down domains cannot form above $300$\,Oe and the local maximum disappears. The actual value of the nucleation field depends on the shape of the sample and defects~\cite{Vial,Zeper}. Although the field above which the local maximum in the FCC curve disappears varies among different samples, the depinning around $130$\,K is a general observation in this compound~\cite{Guguchia,Kassem_M,kassemunconventional,Q_ZhangPRL,Soh,shin,Hu,Shen}. While domain-wall pinning upon cooling is very common in many systems, depinning upon cooling is anomalous. The origin of the depinning, however, is still under investigation. It is possible that the depinning is enabled by the recently observed local structure distortion~\cite{Q_Zhang} which can reduce the local anisotropy energy and thus allows domains to flip, causing the depinning around $130$\,K. 

Since the sample is single domain at high fields, one can wonder if there is still evidence for the magnetic anomaly in the hysteresis loops up to 7\,T. As shown in Fig.~\ref{fig:hysteresis}, the spontaneous magnetization, $M_\mathrm{s}$, is about 0.3\,$\mu_{B}$/Co at $5$\,K~\cite{Wang,Vaqueiro}. We also found that increasing temperature causes the shape of the hysteresis loop to change from a rectangular shape (5\,K), to a triangular-tailed rectangle (100\,K), to a bi-triangular shape (132\,K), until the hysteresis disappears completely (160\,K). The change from triangular-tailed-rectangle to rectangular shape of hysteresis loops can be explained with a domain-wall motion model. As shown in Fig.~\ref{fig:Hnuc_T}, the magnitude of the nucleation field increases as cooling down, until, around $85$\,K, it exceeds the magnitude of the demagnetization field, causing the hysteresis loop changes from a triangular-tailed shape to a rectangular shape~\cite{Tsymbal}. While there is no minimum in the nucleation field, we find that the nucleation field changes its sign around $130$\,K, causing the hysteresis loop changes from a bi-triangular shape to a triangular-tailed shape.

\subsection{Magnetic domain structures from 140~K to 120~K}

Except for a recent MOKE study reported by Lee and coworkers ~\cite{Lee}, domain structures of Co$_3$Sn$_2$S$_2$(0001) single crystal samples are largely unknown. By measuring spatially resolved magnetization through the Kerr effect as a function of temperature, we can examine whether the magnetic anomaly is intrinsic to the physics in a single domain, or, as we suggested previously, is due to the presence of multiple domains and domain walls. Such detailed information is elusive from bulk measurements averaged over the entire sample.

\begin{figure}[h]
\centering
    \includegraphics[width=0.48\textwidth]{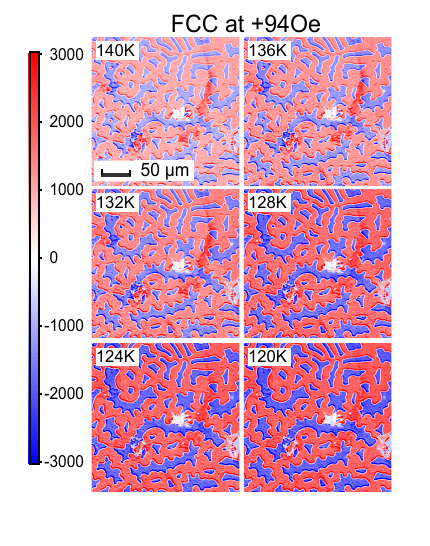}
    \caption{Polar Kerr rotation images of Co$_3$Sn$_2$S$_2$(0001) after FCC in $H_\parallel$ = + $94$\,Oe (pointing out of paper along the c-axis) from room temperature acquired at (a) 140\,K; (b) 136\,K; (c) 132\,K; (d) 128\,K; (e) 124\,K; (f) 120\,K. Red regions are domains with magnetization pointing out of paper along the direction of the applied magnetic field; blue regions are domains with magnetization pointing in the opposite direction. The unit of Kerr rotation is $\mu$rad. The scale bar is 50 $\mu$m. The image size is 240 $\mu$m x 240 $\mu$m.}
    \label{MOKE 130}
\end{figure}

To examine the domain structures near $T = 130$\,K, we acquired a set of MOKE images at $140$\,K, $136$\,K, $132$\,K, $128$\,K, $124$\,K, and $120$\,K after the sample is field cooled in a magnetic field of H$_\parallel=+94$\,Oe. These images are shown in Fig.~\ref{MOKE 130}. Other than the disappearance of a small embedded red domain in the upper left corner (between $136$\,K and $132$\,K) and that of an embedded blue domain to its immediate right (between $128$\,K and $124$\,K), the domain structure remains essentially unchanged between $140$\,K and $120$\,K. Limited by the spatial resolution of the scanning microscope ($0.85\,\mu$m), from these MOKE images alone, we cannot rule out transitions that involve domain walls, such as those proposed in~\cite{Lee}.  

\begin{figure}[h]
\centering
   \includegraphics[width=0.48\textwidth]{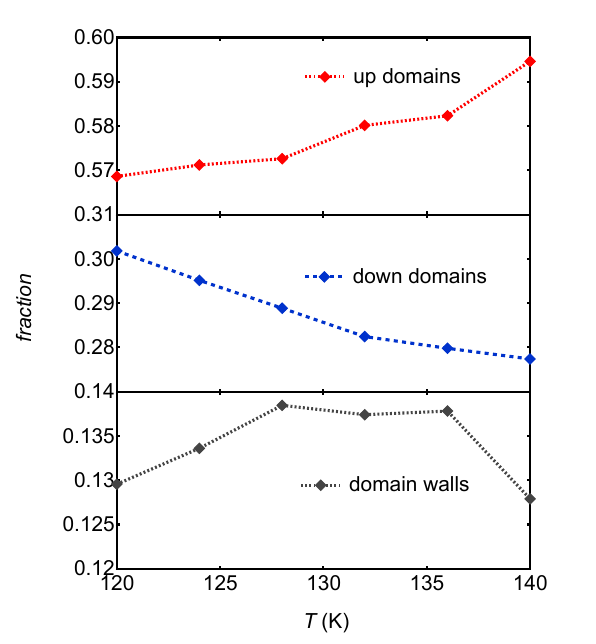}
    \caption{Evolution of the fraction of up and down domains and domain walls while cooling down extracted from MOKE images. The fraction of up domains shrinks while the fraction of down domains grows.}  
    \label{fig:up_down_wall_T}
\end{figure}

We extract the fraction of up domains, down domains, and domain walls from these images as follows. First, we measure the average absolute Kerr rotation of up and down domains. Then, we count points above $75\%$ of the average value as up or down domains, and points below that value as domain walls.  Fig.~\ref{fig:up_down_wall_T} shows the evolution of these fractions as a function of temperature. As the sample is cooled down, the fraction of up domains shrinks while that of down domains grows due to the demagnetization effect. \modif{The fact that the temperature variation of the domain fractions follows the same trend as the evolution expected from the case with no pinning confirms that the magnetic domain walls are indeed able to move between 120K and 140K.}

\section {Conclusion}

We find that the hysteresis between FCC and FCW magnetization in low fields is due to a domain-wall pinning effect and attribute the magnetic anomaly $T_\mathrm{A}$ near 130\,K in the FCC magnetization curve in Co$_{3}$Sn$_{2}$S$_{2}$ to an anomalous domain wall depinning effect upon cooling. We also find that the nucleation field changes its sign around the anomaly temperature $T_\mathrm{A}=130$\,K, causing the shape of the hysteresis loops to change from triangular-tailed rectangles into double-triangles. The MOKE images also reveal a magnetic domain walls depinning between $140$\,K and $120$\,K upon cooling. Our findings corroborate some other domain studies in this compound~\cite{Sugawara,Howlader}. However, the origin the domain-wall depinning still requires further investigation.

\section {Acknowledgments}

V.T., Z.S. and R.R.U. acknowledge support from the UC Lab Fees Research Program (LFR-20-653926) and UC Davis Startup funds. X.D.Z. acknowledges a Visiting Lecture Professorship from Fudan University in support of this work. We acknowledge support from the Physics Liquid Helium Laboratory fund.

\section{Reference}
\bibliographystyle{iopart-num.bst}
\bibliography{Co3Sn2S2}

\end{document}